\documentclass{iaus}
\usepackage{graphicx}

\title[Progenitor of a Ia with short delay?] 
{The progenitor of a type Ia supernova with a short delay time?}

\author[S. Mereghetti et al.]   
{S. Mereghetti$^1$,
N. La Palombara$^1$,
A. Tiengo$^{1,2}$,
P. Esposito$^3$,\\
L. Stella$^4$,
\and G.L. Israel$^4$
}

\affiliation{$^1$  INAF, IASF-Milano,  v. E.Bassini 15, I-20133 Milano, Italy, email: {\tt sandro@iasf-milano.inaf.it} \\[\affilskip]
$^2$IUSS,
v.le Lungo Ticino Sforza 56, I-27100 Pavia, Italy  \\[\affilskip]
$^3$INAF - Oss. Astron. di Cagliari,  loc. Poggio dei Pini, strada 54,
 I-09012 Capoterra, Italy    \\[\affilskip]
$^4$ INAF - Osservatorio Astronomico di Roma, v. Frascati 33, I-00040 Monteporzio Catone, Italy \\[\affilskip]
}

\pubyear{2011}
\volume{281}  
\pagerange{101-104}
\jname{Binary Paths to Type Ia Supernovae Explosions}
\editors{R. Di Stefano  \&  M. Orio, eds.}

\def\approxgt{\mathrel{\hbox{\rlap{\lower.55ex \hbox {$\sim$}}
        \kern-.3em \raise.4ex \hbox{$>$}}}}
\def\approxlt{\mathrel{\hbox{\rlap{\lower.55ex \hbox {$\sim$}}
        \kern-.3em \raise.4ex \hbox{$<$}}}}
\def \xmm {\emph{XMM-Newton} }

\def\ltsima{$\; \buildrel < \over \sim \;$}
\def\lsim{\lower.5ex\hbox{\ltsima}}
\def\gtsima{$\; \buildrel > \over \sim \;$}
\def\gsim{\lower.5ex\hbox{\gtsima}}
\def\msole{~M_{\odot}}
\def\msun{~M_{\odot}}
\def\mdot {\dot M}

\def\rsun{~R_{\odot}}
\def\lsun{~L_{\odot}}

\def\hd {HD\,49798}
\def\rx {RX\,J0648.0--4418}
\def\rxj {RX\,J0648.0--4418}
\newcommand{\src}{HD\,49798/RX\,J0648.0--4418}

\newcommand{\apj}{{\it ApJ}}

\newcommand{\apjs}{{\it ApJS}}
\newcommand{\apjl}{{\it ApJ}}
\newcommand{\aap}{{\it A\&A}}

\newcommand{\mnras}{{\it MNRAS}}
\newcommand{\araa}{{\it ARA\&A}}

\newcommand{\pasp}{{\it PASP}}
\newcommand{\apss}{{\it Ap\&SS}}

\begin{document}

\maketitle

\begin{abstract}

\src\ is the only known X-ray binary composed of a hot subdwarf
and a massive white dwarf (M=1.28$\pm$0.05 $\msole$). This
system, with an orbital period of 1.55 days, is the outcome of
a common envelope evolution, most likely of a pair of stars
with initial masses of $\sim$8--10 $\msun$. When the hot
subdwarf, currently in a He-burning phase, will expand again
and fill its Roche-lobe, the enhanced mass transfer can rapidly
bring the already massive white dwarf above the Chandrasekhar
limit. The possible final fate, either a Type Ia supernova
explosion or an accretion induced collapse, is particularly
interesting in view of the high rotational velocity of this
star, which has the shortest spin period (13 s) observed in a
white dwarf. \keywords{Stars:  subdwarfs, white dwarfs,
individual (\hd); X-rays: binaries, individual (\rxj ).}
\end{abstract}

\firstsection 
\section{Introduction}

The binary \src\ has been  proposed by the organizers of this
conference as one of the ``\textit{mystery object of the
day}''. Indeed, the peculiar properties of the bright star \hd\
attracted the attention of many astronomers since the time of
its first observations in the sixties. The discovery of pulsed
soft X-rays with \textsl{ROSAT} in 1996 renewed interest in
this binary, while our more recent results showed its relevance
for the topics discussed in this IAU Symposium.

Here, after briefly reviewing how the main mystery, i.e. the
nature of the unseen companion of \hd , was finally solved
thanks to X-ray observations, we discuss a few other puzzling
issues still to be clarified to fully understand this unique
binary system.



\section{The peculiar properties of \hd\ }

The first spectroscopic observations of this bright blue star
(B=8, B$-$V=$-$0.27) showed a dominance of He and N lines and
radial velocity variations, pointing to a binary nature.
\cite{jas63} included \hd\  in the (then small) group of early
type subdwarfs and, pointing out the peculiarities of its
spectrum,  concluded their paper with the sentence
``\textit{The orbit of this object would obviously be very
desirable}''. A few years later the orbital period (1.5477
days) and mass function were derived by \cite{tha70}, who also
pointed out the lack of eclipse and/or ellipsoidal variations
in the optical light curve. On the basis of these findings he
suggested that the companion of \hd\ could be a white dwarf,
but the possibility of a late type main sequence star,
outshined in the optical/UV spectra by the much brighter
emission from the hot and luminous subdwarf (L$\sim$10$^4$
$\lsun$), could not be excluded. In the following years several
studies concentrated on a detailed modeling of the star's
atmosphere (e.g., \cite{ric71,duf72}). \hd\ was classified as a
subdwarf of spectral type O6,  with an effective temperature
T$_{eff}$ = 47500$\pm$2000 K, and a surface gravity log\emph{g}
= 4.25$\pm$0.2 \cite{kud78}. The overabundance of Helium,
equalling H in number, was confirmed (mass fractions of
X$_{H}$=0.29 and X$_{He}$=0.78). The low H abundance suggests
that \hd\ is the stripped core of an initially much more
massive and larger star. Also the high abundance of N and low
abundance of C and O confirm that its present surface layers
once belonged to the outer part of the hydrogen-burning core of
a massive star, consistent with an evolution involving a common
envelope phase. The optical mass function could be measured
with great precision by \cite{sti94} (f$_{OPT}$=0.263$\pm$0.004
$\msun$), but all the attempts to reveal the invisible
companion of this single-lined spectroscopic binary were
unsuccessful.

\section{X-rays reveal the nature of the companion star}

The mystery was partially solved in 1996, thanks to the
\textsl{ROSAT} discovery of periodic pulsations at 13 s in the
soft  X-ray (0.1-2 keV) flux \cite{isr97}. This clearly
pointed to the presence of a collapsed object, but it was
impossible to distinguish between a neutron star or a white
dwarf. The very soft spectrum was well fit with a blackbody
plus power-law model, but with a poorly constrained luminosity
in the range from $\sim$10$^{32}$ erg s$^{-1}$ up to more than
10$^{35}$ erg s$^{-1}$ (for the well known distance of 650 pc),
compatible with both possibilities. The X-ray emission from
\hd\ is too soft and faint for satellites like
\textsl{BeppoSAX}, \textsl{ASCA} or \textsl{RossiXTE}, so we
had to wait for \xmm\ in order to make further progress.

After four short observations done in 2002 which added little
information \cite{tie04}, we  managed to obtain a long pointing
of \hd\ in May 2008, strategically scheduled at the time of
conjunction\footnote{The phase of the expected eclipse, that
was never covered in previous X-ray observations.}. Our main
objectives were to exploit the regular X-ray periodicity, which
makes this system equivalent to a double spectroscopic binary,
to constrain the masses of the two stars, and to get a better
estimate of  the source spectral parameters and luminosity. The
measurement of the X-ray pulse delays induced by the orbital
motion, and the discovery of an X-ray eclipse lasting $\sim$1.3
hours, allowed us to derive the X-ray mass function as well as
the system's inclination (79$^{\circ}<i<84^{\circ}$). This
information, coupled to the already known optical mass
function, gives the masses of the two stars: M$_{sd}$ =
1.50$\pm$0.05 $\msun$ for the subdwarf \ and M$_{WD}$ =
1.28$\pm$0.05 $\msun$ for its companion \cite{mer09}.
Furthermore, the high quality spectrum obtained with the
\xmm\/EPIC instrument showed that the total luminosity is only
$\sim$10$^{32}$ erg s$^{-1}$, much smaller than that expected
from a neutron star accreting in the stellar wind of \hd .

\section{More mysteries ?}

While the mysterious companion of \hd\ has finally been
unveiled by X-ray observations, this binary certainly deserves
further study. It stands out in the zoo of accreting X-ray
binaries for being the only one composed of a white dwarf and a
subdwarf. Furthermore, the white dwarf, besides being one of
the most massive white dwarfs with a dynamical mass
measurement, is the one with the shortest spin period (P=13 s).
These properties make \src\ particularly attractive in the
context of type Ia SNe progenitors.

\subsection{Origin of the fast rotation}

Why is the white dwarf rotating so rapidly? Was it born with a
short period or this is the result of spin-up caused by disk
accretion, as in recycled millisecond pulsars?

The radius of \hd\ (1.45$\pm$0.25 $\rsun$) is much smaller than
that of its Roche-lobe. On the other hand this is one of the
few O-type subdwarfs with  evidence for  a relatively strong
stellar wind, with a  mass loss of $\sim$3$\times$10$^{-9}$
$\msun$ yr$^{-1}$  \cite{ham10}. Thus the white dwarf is
accreting through capture of the subdwarf's wind, and there is
no evidence for an accretion disk. If significant spin-up
occurred, this must have been either during the common envelope
phase, which seems unlikely owing to its short duration and
complicated dynamics of the envelope mass, or before it,
possibly when the expanding \hd\ was close to fill its
Roche-lobe. Note that it is relatively easy to spin-up a weakly
magnetized white dwarf, due to the high specific angular
momentum carried by the accreting mass if the accretion disk
extends down to the star's surface \cite{liv98}. If the value
of P=13.2 s observed now has remained close to the equilibrium
period, we can relate the accretion rate during the spin-up
phase, $\mdot$, to the strength of the magnetic field:
$\mdot\sim3\times10^{-10} \mu^2_{30}$ $\msun$ yr$^{-1}$, where
$\mu_{30}$ is the magnetic dipole moment in units of 10$^{30}$
G cm$^3$. If indeed the magnetic field of \rxj\ is as small as
discussed below, one wonders why the white dwarf is not
rotating even faster than observed. In any case, independent of
its origin, the high rotational velocity might have important
consequences for the future evolution of this white dwarf.

\subsection{Properties of the X-ray emission and magnetic field intensity}

The X-ray emission of \rxj\ consists of a soft blackbody
component, with kT=40 eV  and  a strong nearly sinusoidal
modulation (pulsed fraction 56\%), plus  a harder component
dominating above $\sim$1 keV.  The latter can be fit equally
well by a power law with photon index $\Gamma$=1.6 or by a
thermal bremsstrahlung with kT=8 keV, and shows a double peaked
pulse profile \cite{mer11}. Such characteristics are quite
similar to those of cataclysmic variables of the polar and
intermediate polar class, despite these systems are very
different for what concerns their mass donor stars. These
similarities suggest that that the X-ray emission properties
depend mainly on the physical conditions near the white dwarf,
with little influence of the large scale accretion scenario
(wind accretion in \rx\ \textsl{wrt} Roche lobe overflow in
cataclysmic variables).

A condition for accretion onto the white dwarf surface is that
the magnetospheric radius, defined by balancing the magnetic
pressure and the ram pressure of the accreting matter, be
smaller than the corotation radius R$_{C}$ = (G M$_{WD}$
P$^{2}$/4$\pi^{2}$)$^{1/3}$.
This allows us to estimate an upper limit of 2$\times10^{29}$ G
cm$^3$ on the magnetic moment of \rx .  An even stronger limit
of $\mu$$<3\times10^{28}$ G cm$^3$ is obtained assuming that
accretion takes place in the sub-sonic propeller regime, in
which R$_M$$<$R$_{C}$, but the matter is too hot to accrete
\cite{dav79}. These limits  correspond  to  fields of the order
of only a few kG at the white dwarf's surface, much smaller
than those of cataclysmic variables. With such a small field it
is probably difficult to channel the accreting plasma. On the
other hand the large pulsed fraction ($\sim$60\%) observed
below 1 keV requires some form of non-isotropic emission. One
possibility is that the  field has a non-dipolar geometry,
given that the above limits refer only to the dipolar
component.

\subsection{Future evolution}

At the end of the current He-burning phase, \hd\ will expand
again and transfer He-rich matter through Roche-lobe overflow
during a semi-detached phase \cite{ibe94}, but the fraction of
mass that is  retained on the white dwarf is rather uncertain.
Recent  computations, performed assuming the mass accumulation
efficiency  that takes into account the wind mass loss
triggered by the He-shell flashes  \cite{kat04}, indicate that
a mass of 1.4 $\msun$ can be reached after only a few 10$^4$
years \cite{wan10}. However, there are other critical factors
which influence the final fate of the white dwarf, such as,
e.g.,  its composition and rotational velocity.

If  \rxj\ is a CO white dwarf, it could be the progenitor of an
over-luminous type Ia supernova, since the fast rotation can
increase the mass stability limit above the value for
non-rotating stars. Massive white dwarfs are expected to have
an ONe composition, but again the high spin might play a role
here, since it can lead to the formation of  CO white dwarfs
even for high masses \cite{dom96}. Being the result of the
evolution of relatively massive stars ($\sim$8--9 $\msun$),
\src\ would seem a type Ia supernova progenitor with a short
delay time. However, the delay time might be considerably
longer if the explosion has to await that the white dwarf spins
down \cite{dis11}.

Alternatively,  if \rx\ is an ONe white dwarf, an accretion
induced collapse might occur leading to the formation of a
neutron star. The high spin rate and low magnetic field make
this white dwarf an ideal progenitor of a millisecond pulsar.
The evolution of systems like \src\ could  be a promising
scenario for the direct formation of  millisecond pulsars, i.e.
one not involving the recycling of old pulsars in accreting low
mass X-ray binaries.

\vspace{-0.2cm}

\acknowledgements

We acknowledge financial contribution from the agreements
ASI-INAF I/032/10/0. PE acknowledges financial support through
grant Sardegna FSE 2007--2013, L.R. 7/2007.

\end{document}